\documentclass{article}
\usepackage{spconf,amsmath,graphicx}
\usepackage{multirow}
\usepackage{amsmath}
\usepackage{stfloats}
\usepackage{booktabs}
\usepackage{multirow}
\usepackage{bbding}
\usepackage{mathrsfs}
\usepackage{xcolor}
\usepackage{enumitem}
\usepackage{setspace}

\title{Robust Audio-Visual Speech Enhancement: Correcting Misassignments in Complex Environments with Advanced Post-Processing}
\name{Wenze Ren$^1$, Kuo-Hsuan Hung$^{1,2}$, Rong Chao$^{1,2}$, YouJin Li$^1$, Hsin-Min Wang$^2$, Yu Tsao$^2$}
\address{$^1$National Taiwan University, $^2$Academia Sinica}
\begin{document}
%
\maketitle
\begin{abstract}
This paper addresses the prevalent issue of incorrect speech output in audio-visual speech enhancement (AVSE) systems, which is often caused by poor video quality and mismatched training and test data. We introduce a post-processing classifier (PPC) to rectify these erroneous outputs, ensuring that the enhanced speech corresponds accurately to the intended speaker. We also adopt a mixup strategy in PPC training to improve its robustness. Experimental results on the AVSE-challenge dataset show that integrating PPC into the AVSE model can significantly improve AVSE performance, and combining PPC with the AVSE model trained with permutation invariant training (PIT) yields the best performance. The proposed method substantially outperforms the baseline model by a large margin. This work highlights the potential for broader applications across various modalities and architectures, providing a promising direction for future research in this field.
\end{abstract}
\section{Introduction}
\label{sec:intro}

Effective communication through spoken language is a cornerstone of human interaction. However, background noise or competing speakers often pose significant challenges. In noisy environments, it becomes increasingly difficult to discern speech, leading to misunderstandings and reduced communication efficiency. Traditional speech enhancement (SE) technology mainly relies on unimodal methods, which utilize the intrinsic characteristics of audio signals to perform denoising and separation tasks aimed at enhancing speech clarity and intelligibility \cite{choi2019phaseawarespeechenhancementdeep, hu2020dccrndeepcomplexconvolution, subakan2021attentionneedspeechseparation}. While these audio-only methods have made substantial progress by introducing deep learning models~\cite{lu2013speech, wang2018supervised, liu2014experiments}, they often encounter limitations in particularly noisy settings where audio signals alone may not be sufficient for effective enhancement.

Researchers have increasingly focused on multimodal approaches that integrate auditory and visual information in speech enhancement and separation tasks. This shift is based on the understanding that visual cues, such as lip movements and facial expressions, can significantly improve speech comprehension by provide additional contextual information. Supplementing auditory information with visual data can reduce the adverse effects of background noise and competing speakers, thereby improving overall communication effectiveness.

A significant breakthrough in SE came with the introduction of the first multimodal deep convolution method in \cite{8323326}. By integrating visual cues, such as facial features and lip movements, to assist in processing audio signals \cite{10446372, 10022646}, the capability and robustness of SE can be significantly improved. This demonstrates the effectiveness of combining visual and audio information in SE systems~\cite{gao2021visualvoiceaudiovisualspeechseparation, afouras2018conversationdeepaudiovisualspeech, chuang2022improvedliteaudiovisualspeech}.
Building on the success of convolutional neural networks (CNNs), researchers have integrated recurrent neural networks (RNNs) into audio-visual speech enhancement (AVSE) systems~\cite{pandey2020dual}. RNNs, particularly long short-term memory networks (LSTMs), excel at processing time-series data, making them ideally suited for modeling the dynamic characteristics of both speech and visual signals~\cite{7965918}. Attention mechanisms further enhance these models by allowing them to focus on the most relevant features in the audio and visual modalities and adjust the importance of different inputs according to their relevance.
The launch of the Transformer model marks another major advancement in the AVSE field~\cite{10094306}. Transformer's powerful self-attention mechanism can capture long-term dependencies and complex relationships between audio and visual features. This capability makes Transformer particularly effective in environments with multiple competing speakers or severe background noise, further improving the robustness and accuracy of SE systems.

Although these models have achieved significant progress in AVSE, several challenges remain. One major issue is the mismatch between training and test datasets, which hinders the ability of these models to generalize to real-world environments. When faced with noise types not present in the training set, performance drops significantly, even with the addition of visual features. Additionally, noise reduction may not be effective against competing speakers when the quality of the visual features is deficient. These limitations can lead to the poor performance of current AVSE systems and restrict the practical applicability of multimodal SE techniques.

To overcome these challenges, we propose a novel post-processing approach to address the mismatch between training and test datasets and the low-quality visual information that affects the denoising of AVSE systems. The proposed approach takes advantage of multimodal fusion and adjusts the model output to better align with the characteristics of the test environment. Our experiments demonstrate that this additional step increases robustness and adaptability, ensuring improved performance in real-world applications.

\section{Propose Method}

\begin{figure}[!t]
    \centering
    {\includegraphics[width=.5\textwidth]{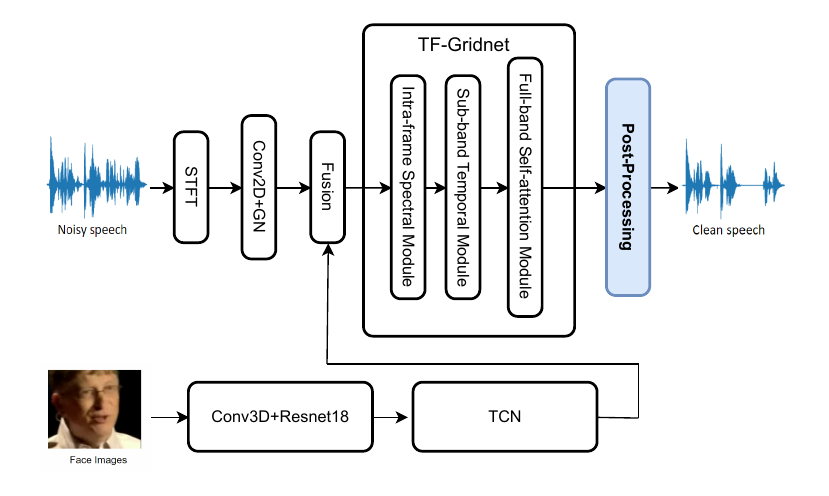}}
    \vspace{-1.0cm}
    \caption{Overview of the proposed audio-visual speech enhancement system.}
    \label{fig:IEMOCAP}
\end{figure}

Our proposed AVSE system consists of two components: the AVSE model and the post-processing classifier (PPC), as shown in Fig.~\ref{fig:IEMOCAP}. The system operates in two stages. First, the AVSE model extracts the target audio from the mixed audio. Then, PPC evaluates both the predicted target audio and the predicted interference audio to select the one with better performance. It is worth noting that the training of PPC is independent of the AVSE model, i.e., PPC does not require the enhanced audio from the AVSE model during training. This design makes PPC a plug-and-play network that can be integrated with various AVSE models. The AVSE model used in this work will be described in Section~\ref{sec:AVSE}, while the PPC used to select better output will be introduced in Section~\ref{sec:PPC}.


\subsection{The AVSE Model}
\label{sec:AVSE}
\subsubsection{Audio Feature Extraction}
The noisy speech signal is initially transformed from the time domain to the time-frequency domain through the Short-Time Fourier Transform (STFT), resulting in a spectral representation. The resulting noisy spectrogram is then processed using convolutional layers combined with global normalization (GN). These layers enhance the representation of the underlying speech signal by delving into deeper levels of the spectrogram to effectively capture salient audio features.

\subsubsection{Visual Feature Extraction}
Visual features are extracted from the input speaker video using a 3D CNN (Conv3D) combined with the ResNet-18 architecture. The Conv3D layers process the temporal and spatial dimensions of the video frames, while ResNet-18 extracts higher-level visual features, resulting in a more robust representation of facial movements. These visual features are then processed using a Temporal Convolutional Network (TCN)~\cite{Afouras_2022}. TCN captures the temporal dependency of the visual features, ensuring more effective modeling of the temporal context of the facial movements.

\subsubsection{Fusion}
We concatenate visual and speech features along the channel dimension, ensuring that the speech features input to TF-Gridnet \cite{wang2023tfgridnetmakingtimefrequencydomain} comprehensively incorporate the speaker's visual information.

\subsubsection{The Separation Network}
We use TF-Gridnet~\cite{wang2023tfgridnetmakingtimefrequencydomain} as the backbone separation network, which is a cutting-edge deep neural network designed for monaural speaker separation, especially under anechoic conditions. Its core functionality revolves around complex spectral mapping, where the network is trained to predict the real and imaginary (RI) components of a target speaker from a mixture signal. This method achieves robust separation by directly modeling phase information, which is often crucial for accurate audio separation.
The architecture of TF-GridNet is composed of multiple blocks, each containing three key modules:
\begin{itemize}[leftmargin=*]
\setlength{\itemsep}{0pt}
\item Intra-frame Spectral Module: This module processes each frame individually, using a bidirectional LSTM (BLSTM) to capture local spectral patterns within each frame. It focuses on modeling frequency-wise dependencies by unfolding the frequency dimension and processing it sequentially.
\item Sub-band Temporal Module: This module handles the temporal dependency across sub-bands by treating each frequency band as a sequence over time. Like the spectral module, it uses a BLSTM to model the temporal dynamics within each frequency sub-band.
\item Full-band Self-attention Module: This module introduces global context by applying self-attention across the entire time-frequency grid. It allows the network to capture long-range dependencies across time and frequency, which is essential for distinguishing overlapping speakers.
\end{itemize}
Together, these modules enable TF-GridNet to effectively exploit local and global spectro-temporal information to achieve state-of-the-art performance in speaker separation tasks. The design of this network enables it to significantly improve the scale-invariant signal-to-distortion ratio (SI-SDR) on benchmark datasets over previous models.

\subsubsection{Loss Functions}
In the SE field, the negative SI-SDR loss is commonly used for model training in most methods, which is defined as:
\begin{equation}
\mathcal{L}_{\text{SI-SDR}}(s, \hat{s}) = -20 \log_{10} \frac{\Vert \frac{\langle \hat{s}, s \rangle}{\Vert s\Vert^2} s \Vert}{\Vert \hat{s} - \frac{\langle \hat{s}, s \rangle}{\Vert s\Vert^2} s \Vert},
\end{equation}
where $s$ and $\hat{s}$ represent the target speech and enhanced speech, respectively.
We also use frequency domain multi-resolution incremental spectral loss \cite{pan2022hybridcontinuitylossreduce}. The total loss function used to train our AVSE model is as follows:
\begin{equation}
\mathcal{L}_{\text{Loss}}(s, \hat{s}) = \mathcal{L}_{\text{SI-SDR}}(s, \hat{s}) + \gamma \frac{1}{M} \sum_{m=1}^{M} \mathcal{L}^{m}_{\text{freq-}\Delta}(s, \hat{s}),
\end{equation}
where \(\mathcal{L}^{m}_{\text{freq-}\Delta}(s, \hat{s})\) represents the spectral loss calculated at a specific resolution. We use three (i.e., $M=3$) configurations for FFT size, hop size, and window length (all in samples), specifically: \{512, 50, 240\}, \{1024, 120, 600\}, and \{2048, 240, 1200\}. The weight $\gamma$ is set to 1.

\begin{figure}[!t]
\centering
\hspace{1cm}
\includegraphics[width=0.5\textwidth]{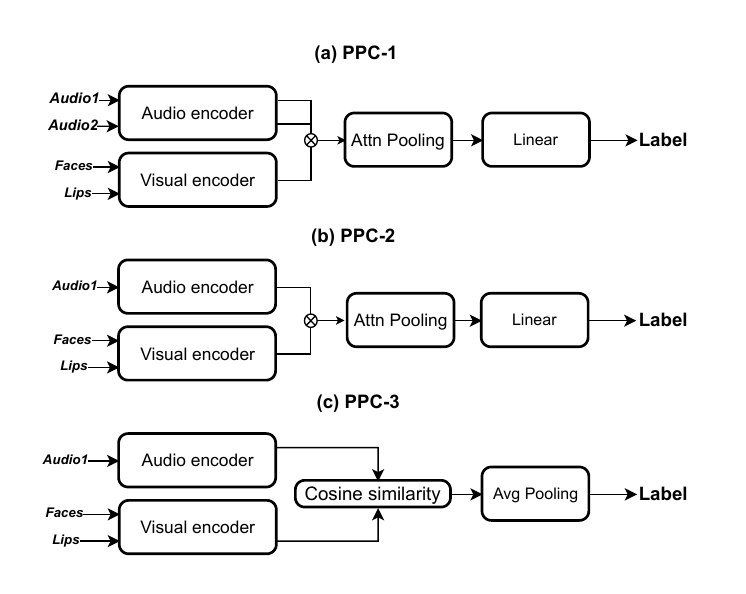}
\vspace{-1.0cm}
\caption{Illustration of three different PPC models. $\otimes$ denotes the concatenation operation.}
\label{fig:system}
\end{figure}

\begin{figure}[!t]
\centering
\hspace*{-0.45cm}
\includegraphics[width=0.55\textwidth]{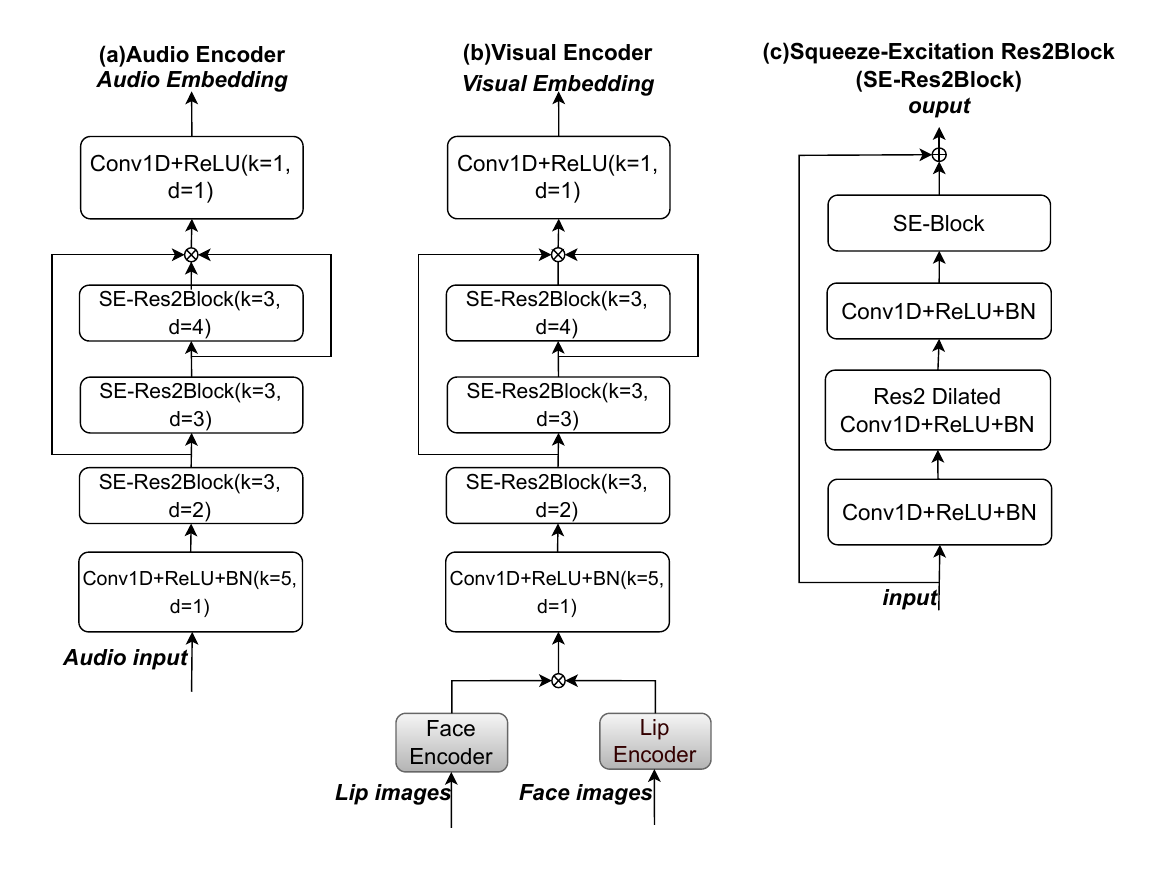}
\vspace{-1.0cm}
\caption{Detailed components of the audio encoder (a) and the visual encoder (b) in the PPC models, and the SE-Res2Block used in both encoders (c). Modules with a gray background are pre-trained and frozen during model training.} 
\label{fig:arc}
\end{figure}

\subsection{The Post-processing Classifier (PPC)}
\label{sec:PPC}
It is observed that AVSE models occasionally output audio that resembles interference audio, such as competing speakers or noise, which degrades model performance. This issue may result from a mismatch between training and evaluation datasets or poor video quality. To address this issue, we introduce PPC, which aims to ensure that the output speech aligns with the target speaker. PPC refines initial model predictions by validating and adjusting output to improve overall accuracy and maintain consistency across datasets. In this study, we propose three distinct PPC models, as shown in Fig.\ref{fig:system}.

\subsubsection{Model Architecture}

The proposed PPC is based on the ECAPA-TDNN \cite{Desplanques_2020} model, which is known for its strong performance in speaker verification. As shown in Fig.~\ref{fig:system}, in PPC, each modality has its encoder to extract the corresponding embeddings. As shown in Fig.~\ref{fig:arc}(a), the audio encoder processes 80-dimensional MFCCs from a 25 ms window with a 10 ms frame shift. As shown in Fig.~\ref{fig:arc}(b), the visual encoder combines face and lip embeddings from a pre-trained face classifier~\cite{Schroff_2015} and a pre-trained lip-reading model \cite{Afouras_2022}, respectively. In this work, we explored different strategies for embedding integration and classifier design, resulting in the following three PPC models:
\begin{itemize}[leftmargin=*]
\setlength{\itemsep}{0pt}
    \item \textbf{PPC-1:} As shown in Fig.~\ref{fig:system}(a), PPC-1 concatenates the embeddings of predicted target speech, predicted interference speech, and visual inputs. The concatenated embeddings are then processed through an attention pooling layer followed by a linear layer to predict the label. PPC-1 takes into account the input order of target and interference audio embeddings. During training, the label is 1 when the order is correct and 0 otherwise.
    \item \textbf{PPC-2:} As shown in Fig.~\ref{fig:system}(b), PPC-2 concatenates the embeddings of predicted target speech and visual embeddings. The concatenated embeddings are then passed through an attention pooling layer followed by a linear layer to predict the label. This setup simplifies the integration of audio and visual features compared to PPC-1. During training, the label is 1 when the audio and visual embeddings are matched (from the same speaker) and 0 otherwise.
    \item \textbf{PPC-3:} As shown in Fig.~\ref{fig:system}(c), this PPC model computes the frame-wise cosine similarity between audio and visual embeddings. The similarity scores are then averaged using a pooling layer to produce the final prediction. During the training phase, the labeling method is the same as PPC-2. Compared to PPC-2, this approach places greater emphasis on the alignment between audio and visual modalities.
\end{itemize}
We used the Binary Cross-Entropy (BCE) loss as the objective function to train our PPC models, which is defined as:
\begin{equation}
\label{eq7}
\begin{aligned}
&\mathcal{L}_{BCE}(y,\hat{y}) = -(y\log(\hat{y})+(1-y)\log(1-\hat{y}),\\
\end{aligned}
\end{equation}
where $y$ and $\hat{y}$ are the ground-truth label and predicted value, respectively. Here, $y$ is 1 when the input audio is the target speech and 0 otherwise.

\subsubsection{Mixup}
We also utilize the mixup \cite{zhang2017mixup} strategy in PPC training, which extends the training distribution by combining the prior knowledge that linear interpolation of target and interference audio should lead to linear interpolation of associated labels. This technique can serve as a form of data augmentation. By incorporating mixup, the model can improve generalization when encountering samples outside the training set.

\subsection{Permutation Invariant Training}

Our investigation found that the low-quality video in the training dataset made it difficult for the AVSE model to use visual features to accurately extract target speech, and that forcing the AVSE model to output target speech under these conditions would actually reduce performance. We solve this problem by applying post-processing to the speech output, allowing the AVSE model to approximate either target speech or interfering speech. We use the permutation invariant training (PIT) \cite{yu2017permutationinvarianttrainingdeep} strategy commonly used in speech separation. The process is formulated as follows:

\begin{equation}
\label{eq2}
\begin{aligned}
s^{\ast} = \mathop{\arg\min}_{\hat{s} \in \{ s_{i},s_{t} \}} \mathcal{L}_{SI-SDR}(s,\hat{s}),
\end{aligned}
\end{equation}
where $s^{\ast}$ is the permutation that minimizes the SI-SDR loss, and $s_{i}$ and $s_{t}$ are the interference speech and target speech, respectively.

\section{Experimental Setup}
\subsection{Dataset}

\subsubsection{Training and validation dataset}
We utilize the officially provided LRS3 dataset (as used in AVSE Challenge 2) as the training and development sets, where noise is pre-mixed. This dataset includes everyday noises and disturbing speakers.  We did not employ dynamic mixing during training.

\subsubsection{Testing dataset}
The proposed PPC models are mainly evaluated from two aspects: 1) the accuracy of the classification task and 2) the SE performance when integrated with the AVSE model. Four test sets were used to evaluate the accuracy of model predictions: the development set in LRS3 (Dev), the evaluation set of AVSE Challenge 2023 (Eva2023), and its two variants Eva2023v1 and Eva2023v2. Eva2023v1 was derived using an AVSE model trained with PIT, while Eva2023v2 was derived using an AVSE model trained without PIT. This setting allows us to compare the performance of PPC models under different training conditions and better understand their robustness. The ground-truth labels 
are obtained as follows:
\begin{equation}
\label{eq8}
\begin{aligned}
&y = 
  \begin{cases}
    1, & \text{if SI-SDR$_{pt}\geq$ SI-SDR$_{pi}$} \\
    0, & \text{otherwise}
  \end{cases}\\
\end{aligned}
\end{equation}
where SI-SDR$_{pt}$ and SI-SDR$_{pi}$ represent the SI-SDR scores of predicted target audio and predicted interference audio, respectively. The predicted interference audio is obtained by subtracting the predicted target audio from the mixed audio.

For evaluating the SE performance of the AVSE model integrated with PPC, we use Eva2023 and the evaluation set of AVSE Challenge 2024 (Eva2024) as the test sets. We use the same objective metrics as the Challenge, including perceptual evaluation of speech quality (PESQ), short-time objective intelligibility (STOI), and scale-invariant signal-to-distortion ratio (SI-SDR), to ensure consistency in the evaluation.



\subsection{Hyperparameters}
For the hyperparameters in the TF-GridNet block~\cite{wang2023tfgridnetmakingtimefrequencydomain}, we set $I=4$, $J=1$, $H=192$, $E=4$, and $L=4$. We performed model training on a single NVIDIA 4090 GPU with a batch size of 2 and an initial learning rate of 0.001. The Adam optimizer was used. If the validation loss did not improve for three consecutive epochs, the learning rate was halved to help the convergence of model training.
Each training epoch took approximately 1.5 hours, and the model typically converged after approximately 60 epochs. Checkpoints were created 
to monitor progress, and early stopping was applied when satisfactory performance was achieved before 60 epochs. 

\begin{table}[!tb]
    \caption{Accuracy of different PPC models on different test sets.}
    \label{table:acc}
    \small
    \vspace*{1.5mm}
    \setlength{\tabcolsep}{3.5pt}
    \centering
    \begin{tabular}{lccccc}
    \toprule[0.4mm]
    \multirow{2}[3]{*}{Model} & \multirow{2}[3]{*}{Mixup} & \multicolumn{4}{c}{Test set} \\
    \cmidrule(lr){3-6}
    &&Dev. & Eva2023 & Eva2003v1 & Eva2003v2 \\
    \midrule[0.4mm]
    PPC-1 & \XSolidBrush & 0.983 & 0.710 & 0.652 & 0.604 \\
    \midrule[0.2mm]
    PPC-2 & \XSolidBrush & 0.997 & 0.828 & 0.764 & 0.712 \\
    \midrule[0.2mm]
    PPC-3 & \XSolidBrush & \textbf{1} & 0.988 & 0.902 & 0.883 \\
    \midrule[0.2mm]
    PPC-3 & \Checkmark & \textbf{1} & \textbf{0.995} & \textbf{0.939} & \textbf{0.890} \\
    \bottomrule[0.4mm]
    \end{tabular}%
\end{table}

\begin{table}[!tb]
    \caption{SE performance of different AVSE models on Eva2023 and Eva2024.}
    \label{table:se}
    \vspace*{1.5mm}
    \setlength{\tabcolsep}{3pt}
    \centering
    \begin{tabular}{lcccccc}
    \toprule[0.4mm]
    \multirow{2}[3]{*}{Test set} & \multirow{2}[3]{*}{Method} & \multicolumn{2}{c}{Setting} & \multicolumn{3}{c}{Evaluation} \\
    \cmidrule(lr){3-4} \cmidrule(lr){5-7}
    && PIT & PPC & PESQ & STOI& SI-SDR \\
    \midrule[0.4mm]
    \multirow{6}*{Eva2023} & Noisy & -- & -- & 1.14 & 0.44 & -5.1 \\
    \cmidrule(lr){2-7}
    & Baseline & -- & -- & 1.41 & 0.55 & 3.67 \\
    \cmidrule(lr){2-7}
    & \multirow{4}*{Ours} & \XSolidBrush & \XSolidBrush
 & 1.89 & 0.78 & 6.35 \\
    \cmidrule(lr){3-7}
    && \XSolidBrush & \Checkmark & 1.89 & 0.81 & 7.30 \\
    \cmidrule(lr){3-7}
    && \Checkmark & \XSolidBrush & 1.82 & 0.77 & 5.41 \\
    \cmidrule(lr){3-7}
    && \Checkmark & \Checkmark & \textbf{1.91} & \textbf{0.82} & \textbf{7.94}\\
    \midrule[0.4mm]
 \multirow{3}*{Eva2024} & Noisy & -- & -- & 1.47 & 0.61 & -5.49 \\
    \cmidrule(lr){2-7}
    & \multirow{2}*{Ours} & \Checkmark & \XSolidBrush
 & 2.17 & 0.72 & 3.80 \\
 \cmidrule(lr){3-7}
    && \Checkmark & \Checkmark & \textbf{2.22} & \textbf{0.77} & \textbf{5.71}\\
    \bottomrule[0.4mm]
    \end{tabular}%
\end{table}

\section{Experimental Results}

\subsection{Accuracy of PPC}

The accuracy of different PPC models on different test sets is shown in Table~\ref{table:acc}. We can see that the PPC models using a single speech input (PPC-2 and PPC-3) outperform the PPC model using two speech inputs simultaneously (PPC-1). 
We speculate that the result is due to the model needing to learn that different orders of the two audio files will produce different outputs, which may be too challenging for the model. Furthermore, the model that integrates speech and video before prediction (PPC-2) performs worse than the model that directly calculates the distance between the two modalities in the latent space (PPC-3). This is mainly because directly calculating the distance between the two modalities allows the visual encoder to learn audio embeddings that are closer to the target speech. In addition, applying the mixup training strategy can further enhance the performance of PPC-3, with the accuracy reaching 99.5\%, 93.9\%, and 89\% on the Eva2023, Eva2023v1, and Eva2023v2 sets, respectively.

\begin{figure}[!tb]
\centerline{\includegraphics[width=0.9\columnwidth]{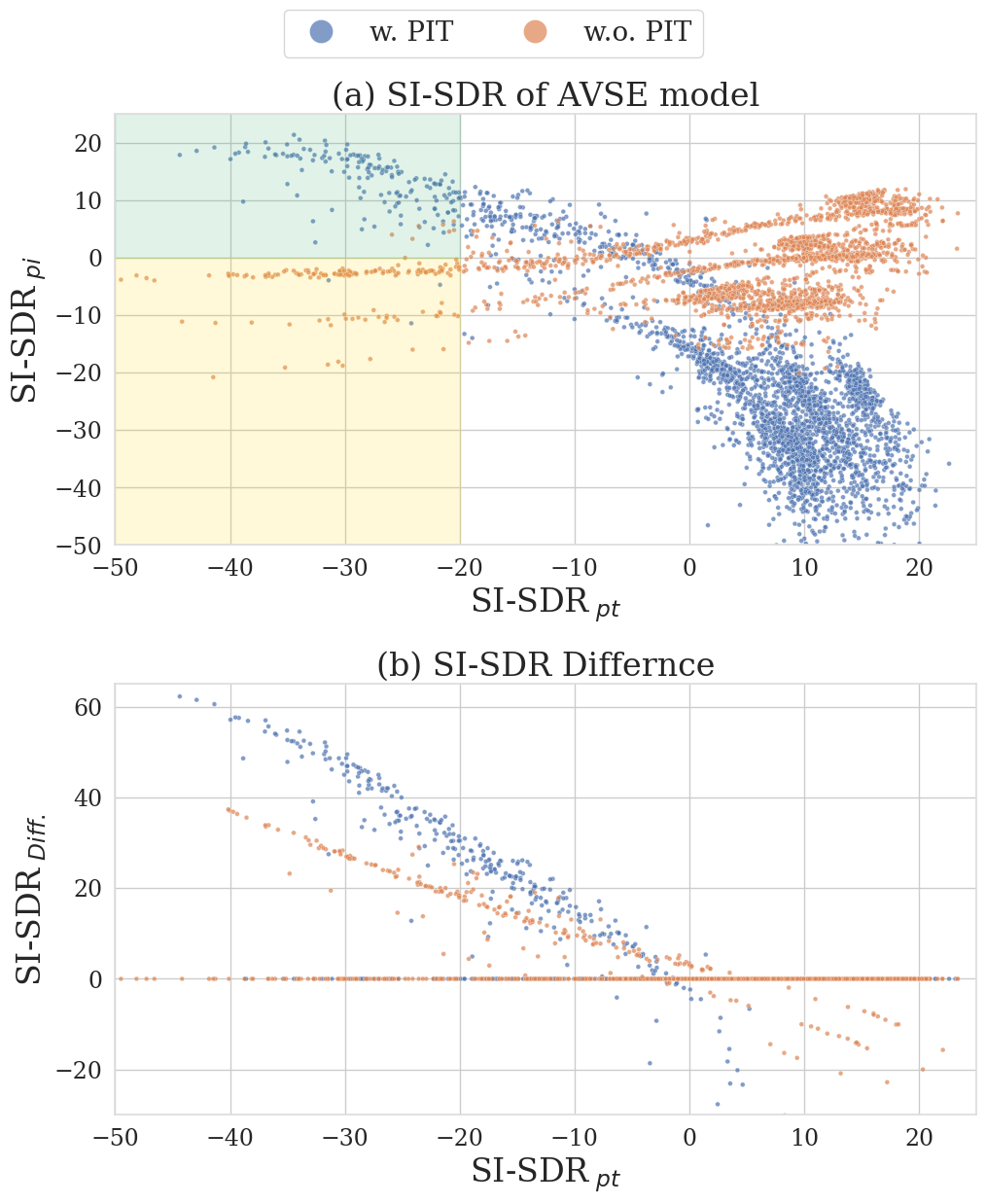}}
\vspace{-0.3cm}
\caption{Scatter plot of SI-SDR scores. (a) SI-SDR scores of predicted target audio (SI-SDR$_{pt}$) and interference audio (SI-SDR$_{pi}$) produced by different AVSE models. (b) Difference in SI-SDR (SI-SDR$_{Diff.}$) before and after applying the best PPC model (PPC-3 with mixup in Table~\ref{table:acc}).}
\label{fig:scatter}
\end{figure}

It is worth noting that the accuracy on Eva2023v1 is higher than that on Eva2023v2.
It is speculated that the AVSE model tends to output audio that resembles interference audio under certain circumstances, especially when the input video quality is poor. Forcing the AVSE model to learn the correct speech may end up outputting a mixture of target audio and interference audio. In contrast, the AVSE model trained with PIT only needs to output audio that is close to either the target audio or interference audio, making it easier for the PPC model to make accurate judgments. This is confirmed by the scatter plot in Fig.~\ref{fig:scatter}(a). In the case where the SI-SDR value of the predicted target audio is low (see left, below -20 dB on the x-axis), for the AVSE model without PIT (orange dots), the corresponding SI-SDR value of the predicted interference audio (y-axis) is mostly below 0 dB (yellow shaded area). In contrast, for the AVSE model trained with PIT (blue dots), the corresponding SI-SDR value of the predicted interference audio (y-axis) is mostly above 0 dB (green shaded area). This result indicates that the AVSE model trained with PIT can be better improved through PPC.

\subsection{AVSE Performance}
Fig.~\ref{fig:scatter}(b) shows the difference in SI-SDR after applying PPC to the AVSE model. A difference greater than 0 indicates that PPC selected a better SE output. It can be clearly seen from the figure that for most audio files, PPC can correctly classify and effectively improve SE performance (SI-SDR$_{Diff.}$ $\geq$ 0 dB). Table~\ref{table:se} shows the objective evaluation scores for the AVSE challenge test sets. The evaluation results on Eva2023 show that regardless of whether the AVSE model is trained with PIT, using PPC can effectively improve the final SE performance, especially in terms of SI-SDR. The combination of PPC and the PIT-trained AVSE model achieved the best results, outperforming the baseline model provided by the 2023 challenge by 0.5 in PESQ, 0.27 in STOI, and 4.27 in SI-SDR. We also submitted our model outputs for the 2024 challenge test set Eva2024. The results show that PPC improves SE performance by 0.05 in PESQ, 0.05 in STOI, and 1.91 in SI-SDR. These results demonstrate the effectiveness of PPC as a post-processing for AVSE models.

\section{Conclusions}
We have observed that various factors in audio-visual speech enhancement can lead to erroneous speech output, such as poor video quality or domain differences between the training and test datasets, which often degrade model performance. This paper proposes the use of a Post-Processing Classifier (PPC) to correct the output speech. With PPC, we do not need to force the AVSE model to output correct speech; instead, we use the Permutation Invariant Training (PIT) strategy to let the AVSE model output either clean speech or noise, thereby achieving a more stable training process. Experimental results show that the combination of PPC and the PIT-trained AVSE model achieves the best SE performance, significantly outperforming the baseline model in the AVSEC challenge. Our findings provide a novel perspective into the field of audio-visual speech enhancement.

There are potential areas for future work. The proposed PPC is not limited to visual input; further experiments are needed to evaluate its effectiveness with other modalities. Moreover, exploring its impact on other audio-visual speech enhancement models is necessary for further advancement.
\setstretch{0.86}
\bibliographystyle{IEEEbib}
\bibliography{strings,refs}

\end{document}